# TSO-DSO Coordination for the Procurement of Balancing and Congestion Management Services: Assessment of a meshed-to-meshed topology

Leandro Lind, Rafael Cossent, and Pablo Frías

*Abstract*— **This paper proposes a comprehensive model for different Coordination Schemes (CSs) for Transmission (TSO) and Distribution System Operators (DSO) in the context of distributed flexibility procurement for balancing and congestion management. The model proposed focuses on the coordination between the EHV (TSO) and the HV (DSO) levels, exploring the meshed-to-meshed topology, including multiple TSO-DSO interface substations. The model is then applied to a realistic case study in which the Swedish power system is modeled for one year, considering a representation of the transmission grid together with the subtransmission grid of Uppsala city. The base case scenario is then subject to different scalability and replication scenarios. The paper corroborates the finding that the Common CS leads to the least overall cost of flexibility procurement. Moreover, it shows the effectiveness of the Local Flexibility Market (LFM) for the DSO in the Swedish context in reducing potential penalties in a Multi-level CS.**

*Index Terms*— **TSO-DSO coordination, flexibility markets, optimization, power system economics, electricity markets.**

## Nomenclature

Indexes:
| | |
|---|---|
| $h \in H$ | hour. |
| $i, j, ii \in N$ | node. |
| $g \in G$ | generator. |
| $f \in F$ | flexibility service provider (FSP). |
| $s \in S$ | system operator (SO). |
| $t \in T$ | system operator type. |
| $z, zz \in Z$ | bidding zones. |
| $lv \in LS$ | levels of subscription of the interface. |

Sets:
| | |
|---|---|
| $L(i,j)$ | Set of lines from node $i$ to node $j$. |
| $TS(t,s)$ | Set of correspondence between $t$ and $s$. |
| $SUBS$ | Set of substations nodes ($SUBS \subset N$). |
| $INTER$ | Set of interface nodes ($INTER \subset N$). |
| $IS(i,s)$ | Set of nodes $i$ belonging to System Operator $s$. |
| $IG(i,g)$ | Set of generators $g$ connected at node $i$. |
| $IF(i,f)$ | Set of FSPs $f$ connected at node $i$. |
| $ZN(i,z)$ | Set of nodes $i$ in bidding zone $z$. |
| $ZG(g,z)$ | Set of generators $g$ in bidding zone $z$. |
| $IZ(z,zz)$ | Set of interconnections between bidding zones. |
| $RES(g)$ | Subset of generators $g$ that are RES. |
| $ESS(f)$ | Subset of FSPs that are ESS. |

Parameters:
| | |
|---|---|
| $Q_g^+$ | Maximum output of generator $g$ in MW. |
| $D_{i,h}$ | Demand at node $i$ in hour $h$ in MW. |
| $Q_{f,h}^+ / Q_{f,h}^-$ | Maximum/minimum output of FSP $f$ in hour $h$ in MW. |
| $X_{i,j}$ | Reactance of line $(i,j)$ in p.u. |
| $P_{i,j}^+ / P_{i,j}^-$ | Max./min. power flow of line $(i,j)$ in MW. |
| $\theta_i^+ / \theta_i^-$ | Maximum/minimum angle $\theta$ for node $i$ in p.u. |
| $Bid_g$ | Bid of gen. $g$ in the DA market in €/MWh. |
| $Bid_f$ | Bid of FSP $f$ in the AS market(s) in €/MWh. |
| $DispatchDA_{i,h}$ | Total generation cleared in the DA market produced in node $i$ during hour $h$ in MW. |
| $QDA_{f,h}$ | Quantity dispatched in the DA for FSP $f$ in hour $h$. |
| $Imb_{g,h}$ | Imbalance of generator $g$ in hour $h$ in MW. |
| $SB$ | Base Power in MW. |
| $Cyc_g$ | Cycling cost of generator $g$ in €/MW. |
| $ResProfile_{g,h}$ | Profile for RES $g$ in hour $h$ in p.u. |
| $MinDispatch_g$ | Minimum technical dispatch of gen. $g$ in p.u. |
| $NTC_{z,zz}^+ / NTC_{z,zz}^-$ | Upper/lower bound for Net Transfer Capacity between bidding zones $z, zz$. |
| $MaxFlex_f^+ / MaxFlex_f^-$ | Maximum upward/downward flexibility capacity in relation to the DA dispatch for RES FSP $f$ in p.u. |
| $SoC_{f,h=1}^{init}$ | Initial SoC of ESS $f$ in hour 1. |
| $EF$ | ESS efficiency. |
| $DRMaxh$ | Number of hours that a DR can provide flex. |
| $MinBidSize$ | Minimum bid size in MW. |
| $Impact_i$ | Average of PTDFs for substation $i$ in p.u. |
| $CNSF$ | Cost of non-served flexibility in €/MWh. |
| $CSubs_{i,lv}$ | Cost per subscription level $lv$ in node $i$ in €/MWh. |
| $DispatchFSP_{f,h}$ | Parameter that captures the activation of FSP $f$ in the local flexibility market. |
| $PTDF_{i,j,ii}$ | PTDF of node $ii$ over line $i,j$. |

L. Lind (corresponding author), R. Cossent, and P. Frías are with Universidad Pontificia Comillas, Escuela Técnica Superior de Ingeniería - ICAI, Instituto de Investigación Tecnológica, Madrid, Spain. E-mails: leandro.lind@iit.comillas.edu, rafael.cossent@iit.comillas.edu, pablo.frias@iit.comillas.edu.

Variables:

| | |
|---|---|
| $p_{i,j,h}$ | Power flow in line connecting nodes $i$ and $j$ during hour $h$ in MW. |
| $\theta_{i,h}$ | Angle θ at node $i$ in hour $h$ in radians. |
| $qda_{g,h}$ | Quantity cleared in the DA market for generator $g$ in hour $h$ in MW. |
| $dda_{i,h}$ | Total generation cleared in the DA market produced in node $i$ during hour $h$ in MW. |
| $q_{f,h}$ / $q_{f,h}^{up}$ / $q_{f,h}^{dw}$ | Quantity cleared in the AS market for FSP $f$ in hour $h$ in absolute, upward and downward values in MW. |
| $psubs_{i,h}$ | Power leaving or entering substation $i$ in hour $h$ ( $i \in SUBS$) in MW. |
| $su_{g,h}$ | Start-up of generator g in hour $h$; $\in \{0,1\}$. |
| $sd_{g,h}$ | Shutdown of generator g in hour $h$; $\in \{0,1\}$. |
| $uc_{g,h}$ | Unit commitment status of generator g in hour $h \in \{0,1\}$. |
| $tc_{z,zz,h}$ | Transfer capacity between bidding zones $z$ and $zz$ in hour $h$ in MW. |
| $qmax_{g,h}$ | Maximum generation of $g$ in hour $h$ within the unit commitment problem in MW. |
| $nsf_{i,h}^{p}$ / $nsf_{i,h}^{n}$ | Non-served upward (p) and downward (n) flexibility in node $i$ in hour $h$ in MW. |
| $soc_{f,h}$ | State of Charge of ESS $f$ in hour $h$ in p.u. |
| $pdis_{f,h}$ / $pcha_{f,h}$ | Power being discharged/charged from ESS $f$ in hour $h$ in p.u. |
| $flexess_{f,h}^{dw}$ | Auxiliary variable for the implementation of the storage logic. |
| $bcha_{f,h}$ / $bdis_{f,h}$ | Binary variable for the charging/discharging of ESS $f$ in hour $h$. |
| $uc_{f,h}^{up}$ / $uc_{f,h}^{dw}$ | Binary variable for the upward/downward activation of FSP $f$ in hour $h$. |
| $vd_{s,h}$ | Virtual demand of DSO $s$ in hour $h$. |
| $pimport_{i,h}$ | Power imported by the distribution grid from the transmission grid at node $i$ in hour $h$ in MW. |
| $pexport_{i,h}$ | Power exported from the transmission grid to the distribution grid at node $i$ in hour $h$ in MW. |
| $psubs_{i,h}$ | Power demanded at the interface at node $i$ in hour $h$ in MW. |
| $qsubs_{i,lv,h}$ | Level $lv$ of use of subscription power at the at node $i$ in hour $h$ in MW. |
| $pfsub_{i,j,h}$ | Impact of node $j$ on the power flow over node $i$ in hour $h$ in MW. |

## I. Introduction

POWER systems have evolved from local decentralized grids in the late 1800s to nationally interconnected systems in which generation is often far from the load, connected by long transmission lines [1]. Modern power systems, however, are seeing a resurgence in distributed energy resources (DER), which are expected to coexist with bulk generation, contributing to the efficiency of the system and to the current decarbonization and electrification targets.

From a network's perspective, the higher share of DERs means that Distribution System Operators (DSOs) and Transmission System Operators (TSOs) will both be able to procure system services from these new actors, increasing the efficiency in grid operation and liquidity in service markets. In fact, several initiatives in Europe have already started being implemented in order to allow for the DSO to procure local flexibility from DER for grid management purposes [2]. From the TSO side, too, existing markets are being opened to the participation of distribution-connected units. As an example, the current European regulation mandates that demand response shall be allowed to participate in balancing markets [3]. However, in a future scenario with more flexibility needs from both TSOs and DSOs, system operators will need to coordinate so that procurement and activation of the flexibility provided by DERs are done in the most efficient way and without compromising the security of the system [4]. In this context, a recent line of research has been devoted to the proposition and analysis of different Coordination Schemes (CSs). CSs can be understood as the relation between TSO and DSO in terms of roles and responsibilities for the procurement and deployment of system services provided by distribution-connected resources [5]. In general terms, the provision of such services by DERs will be made through the exploitation of their explicit flexibility, defined as the ability to modify generation and/or consumption patterns in accordance with a commercial agreement, often traded in organized markets [6], [7].

The CSs proposed in the literature focus on how roles and responsibilities are allocated to TSOs, DSOs and Market Operators (MOs). The authors in [5], [8] propose five different CSs, namely centralized, common, local, integrated and shared responsibility CSs. Several other papers have also proposed CSs, as reviewed in [4], [9]. Outside the academic literature too, CSs have been proposed and analyzed. Several research projects have demonstrated different CSs in multiple European countries [10], [11].

Additionally, the European associations representing TSOs and DSOs have published a joint paper in which three CSs are proposed for the procurement of balancing and congestion management services [12]. In this report, option 1 proposes separate congestion management markets for TSO and DSO plus a balancing market operated by the TSO. Option 2 is a combined TSO and DSO congestion management market and a separated balancing market. Option 3 consists of a combined balancing and congestion management market for both TSO and DSO, also considered in [13].

In the analysis of the proposed CSs, several works have conducted a qualitative assessment of the different CSs. The authors in [14] analyzed common, multi-level and fragmented market models in both joint and disjoint configurations, concluding that a common market model leads to the least cost of procurement, which corroborates the findings of previous works [15]. However, other CSs could converge to the efficiency of the common market if the power flow at the TSO-DSO interface is correctly priced. Another challenge of the common CS is the allocation of flexibility procurement costs to TSO and DSO. The authors in [16] propose several methods for this allocation, along with their properties.

The body of literature analyzing the techno-economic properties of CSs, however, has focused on the theoretical development of CS algorithms and their testing on stylized or test case networks. Another characteristic of these works is the generalization of a meshed transmission grid and a radial distribution grid. On the European TSO-DSO landscape, however, these characteristics are not always observed. According to [17], only in 6 EU Member States, the DSO operates LV and MV assets only, and in 19 Member States, DSOs operate the 110kV or higher voltage levels. In this context, TSO-DSO coordination is also expected to happen in meshed-to-meshed topologies. One important difference in the modelling of CSs, in this case, is the existence of multiple TSO-DSO interfaces.

This paper aims to provide a comprehensive TSO-DSO model, including different CSs focused on the TSO (EHV) and the HV DSO grids. This model is then applied to the Swedish power system considering real consumption and generation data for one year. Finally, several scalability and replicability scenarios are analyzed.

More specifically, the contributions of this paper are summarized as follows:
1) The modelling of different CSs for the procurement of balancing and congestion management in meshed-meshed grids with multiple TSO-DSO interfaces.
2) The application of different coordination to a realistic system and case study.
3) The testing of the behaviour of coordination schemes for different scenarios of scalability and replicability.

The rest of this paper is organized as follows: In Section II, the modelling of the different CSs is presented. Section III presents the Swedish case study. Section IV presents the results and Section V concludes.

## II. Systems and Coordination Schemes

In this paper, a combined TSO-DSO model for the procurement of flexibility, including flexibility from DERs, is presented. This model aims to provide an assessment tool for different TSO-DSO coordination schemes, considering that both System Operators (SOs) may utilize resources connected to the distribution grid. Therefore, the model proposed is composed of three building blocks, namely, a wholesale energy market, a congestion management market, and a balancing market.

We model what aims to describe a generic European market sequence, as described in [12], [18]. This means that firstly, a Day-Ahead (DA) wholesale energy market is operated without considering network constraints other than between bidding zones. The DA is followed by a congestion management market in which the system operator (TSO and DSO, depending on the CS) will check for the feasibility of the dispatch. Finally, a balancing market is run to adjust mismatches between what was scheduled in the DA and what is delivered in real-time.

Four main CSs are analyzed in this paper, namely the Common CS, and the Multi-level CS, both in a joint balancing-congestion management configuration and a separate one. The Common CS is defined as a single market in which both congestions and imbalances are solved (jointly or in sequence). It can be assumed that this market is run by a single entity (e.g. an independent market operator or the TSO) and that both TSO and DSO procure flexibility in this market. However, it is not modelled how TSOs and DSOs will share the cost of the market clearing, as proposed in [16].

The Multi-level CS considers that firstly, the DSO is responsible to run a local congestion management market to solve congestions at the distribution grid, followed by the TSOs markets. In this sequence of markets, unused bids by the DSO are then passed on to the TSO market(s), if they do not create additional constraints.

For each CS, two variations exist, namely joint and separate. Fig. 1 presents an overview of the modelling approach described. Each blue box represents one market session, analyzed through Mixed Integer Programming (MIP) optimization problems. The models are implemented in General Algebraic Modeling System (GAMS) and described in what follows.

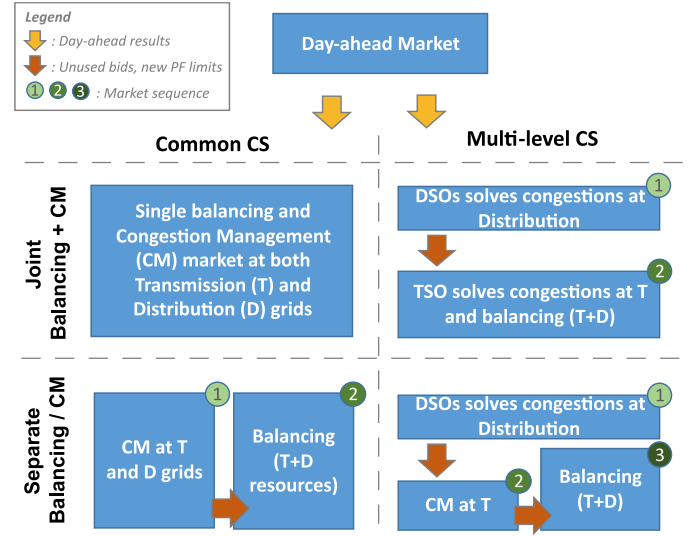

**Fig. 1.** Overview of the four CSs analyzed.

### A. Day-Ahead Market

The day-ahead market is characterized by a clearing of the total demand in each hour and the merit order list of generation bids. At this market phase, the network is not taken into account, except for the limits between bidding zones. Therefore, the MO minimizes the generation cost (1), ensuring that the total demand within the bidding zone is supplied while accounting for imports from and exports to other bidding zones (2). Equation (3) computes the day-ahead dispatch that will later be passed on to the following CSs. Equations (4)-(6) are responsible for the unit commitment logic (based on [19]). Equation (7) accounts for any minimum uptime if applicable to the generator g (e.g. nuclear or thermal power plants), while (8) accounts for the minimum technical dispatch. Finally, (10) and (11) limit transfer capacity between bidding zones and power output per generator to their bounds.

$$\min \sum_{g,h} [(Bid_g \cdot qda_{g,h}) + (su_{g,h} \cdot Cyc_g \cdot Q_g^+)] \quad (1)$$

s.t.

$$\sum_{g \in ZG} qda_{g,h} - \sum_{zz \in IZ} tc_{z,zz,h} + \sum_{zz \in IZ} tc_{zz,z,h} = \sum_{i \in ZN} D_{i,h} : (\lambda_{z,h}) \quad \forall z, h \quad (2)$$

$$dda_{i,h} = \sum_{g \in IG} qda_{g,h} \quad \forall i, h \quad (3)$$

$$uc_{g,h-1} = uc_{g,h} - su_{g,h} + sd_{g,h} \quad \forall g, h \quad (4)$$

$$qmax_{g,h} = uc_{g,h} \cdot Q_g^+ \quad \forall g, h \quad (5)$$

$$qda_{g,h} \leq qmax_{g,h} \quad \forall g, h \quad (6)$$

$$su_{g,h} \cdot minup_g \leq \sum_{h}^{h+minup_g} uc_{g,h} \quad \forall g, h \quad (7)$$

$$uc_{g,h} \cdot Q_g^+ \cdot MinDispatch_g \leq qda_{g,h} \quad \forall g, h \quad (8)$$

$$q_{g,h} \leq Q_g^+ \cdot ResProfile_{g,h} \quad \forall h, g \in RES \quad (9)$$

$$NTC_{z,zz}^- < tc_{z,zz,h} < NTC_{z,zz}^+ \quad \forall z, zz, h \quad (10)$$

$$qda_{g,h} < Q_g^+ \quad \forall g, h \quad (11)$$

Following the DA market, the system service market(s) take place based on the results of the DA market. Hence, the optimal dispatch for all generators is passed on as a parameter to all subsequent service markets in all CSs. The dispatch is passed on aggregated per node (12).

$$DispatchDA_{i,h} = dda_{i,h}^* \quad \forall i, h \quad (12)$$

### B. Common Joint CS

In this Common Joint CS, the market solves all imbalances and network congestions using resources connected to both the transmission and the distribution networks. In this case, a single minimization problem is solved. Equation (13) minimizes the total cost of the Flexibility Service Providers' (FSP) activation during 24h. Considering that both congestion management and balancing markets are centrally run, the modelling of this CS can be seen as a single DC Optimal Power Flow (OPF). However, the demand balance equations (14)-(15) and the power flow equations are split according to the type of SO (16)-(17) (based on [15], [20]). In addition, (18) and (19) ensure that the power at the substation (the interface between TSO and DSO) is consistent. Equations (20)-(21) limit the maximum number of hours (in equivalence of power output) that DR FSPs can provide flexibility in order to account for comfort limitations in flexibility provision. Equations (22)-(23) limit the upward and downward provision of flexibility from RES in relation to their DA dispatch, given that these types of FSP may only have a limited capacity upward, especially (e.g. due to forecasting errors). Equations (24)-(30) are an implementation for the Energy Storage System (ESS) logic (based on [21]).

Equations (31)-(35) implement the unit commitment equivalent for the FSPs. Finally, (36)-(39) limit the power flow over the lines, the angles in each node, and the maximum output per FSP.

$$\min \sum_{s \in TS \wedge t=TSO, i, f, h} \left[(Bid_f \cdot q_{f,h}^{up}) + (Bid_f \cdot q_{f,h}^{dw})\right]$$
$$+ \sum_{s \in TS \wedge t=DSO, i, f, h} \left[(Bid_f \cdot q_{f,h}^{up}) + (Bid_f \cdot q_{f,h}^{dw})\right] \quad (13)$$
$$+ \sum_{i,h} \left[(nsf_{i,h}^{up} + nsf_{i,h}^{dw}) \cdot CNSF\right]$$

s.t.

$$DispatchDA_{i,h} + \sum_{f \in IF} q_{f,h} - \sum_{j} p_{i,j,h} + \sum_{j} p_{j,i,h} - Imb_{i,h}$$
$$= D_{i,h} + nsf_{i,h}^p - nsf_{i,h}^n \quad \forall i \in IS, (s \in TS) \wedge (t = TSO), h \quad (14)$$

$$DispatchDA_{i,h} + \sum_{f \in IF} q_{f,h} - \sum_{j} p_{i,j,h} + \sum_{j} p_{j,i,h} - Imb_{i,h}$$
$$= D_{i,h} + nsf_{i,h}^p - nsf_{i,h}^n \quad \forall i \in IS, (s \in TS) \wedge (t = DSO), h \quad (15)$$

$$p_{i,j,h} = SB \cdot \frac{\theta_{i,h} - \theta_{j,h}}{X_{i,j}} \quad \forall (i \in IS, j \in IS) \in L, (s \in TS) \wedge (t = TSO), h \quad (16)$$

$$p_{i,j,h} = SB \cdot \frac{\theta_{i,h} - v\theta_{j,h}}{X_{i,j}} \quad \forall (i \in IS, j \in IS) \in L, (s \in TS) \wedge (t = DSO), h \quad (17)$$

$$p_{i,j,h} = SB \cdot \frac{v\theta_{i,h} - v\theta_{j,h}}{X_{i,j}} \quad \forall (i,j) \in L \wedge [(i \in SUBS) \vee (j \in SUBS)], h \quad (18)$$

$$\sum_{j} p_{j,i,h} = \sum_{j} p_{i,j,h} \quad \forall (i,j) \in SUBS, h \quad (19)$$

$$\sum_{h} q_{f,h}^{up} \leq \sum_{h} Q_{f,h}^+ \cdot DRMaxh \quad \forall f \in DR \quad (20)$$

$$\sum_{h} q_{f,h}^{dw} \leq \sum_{h} Q_{f,h}^- \cdot DRMaxh \quad \forall f \in DR \quad (21)$$

$$q_{f,h}^{up} \leq QDA_{f,h} \cdot MaxFlex_f^+ \quad \forall f \in RES, h \quad (22)$$

$$q_{f,h}^{dw} \leq QDA_{f,h} \cdot MaxFlex_f^- \quad \forall f \in RES, h \quad (23)$$

$$soc_{f,h} = SoC_{f,h=1}^{init} + soc_{f,h-1} - (pdis_{f,h} \cdot Q_{f,h}^+ \cdot EF + q_{f,h}^{up}) + (pcha_{f,h} \cdot Q_{f,h}^- \cdot EF + q_{f,h}^{dw}) \quad \forall f \in ESS, h \quad (24)$$

$$soc_f^- \leq soc_{f,h} \leq soc_f^+ \quad \forall f \in ESS, h \quad (25)$$

$$q_{f,h}^{dw} = flexess_{f,h}^{dw} \cdot Q_{f,h}^+ \cdot EF \quad \forall f \in ESS, h \quad (26)$$

$$q_{f,h}^{up} = flexess_{f,h}^{up} \cdot Q_{f,h}^+ \cdot EF \quad \forall f \in ESS, h \quad (27)$$

$$bcha_{f,h} + bdis_{f,h} \leq 1 \quad \forall f \in ESS, h \tag{28}$$

$$pdis_{f,h} + flexess_{f,h}^{dw} \leq bdis_{f,h} \quad \forall f \in ESS, h \tag{29}$$

$$pcha_{f,h} + flexess_{f,h}^{up} \leq bcha_{f,h} \quad \forall f \in ESS, h \tag{30}$$

$$q_{f,h}^{up} \leq Q_{f,h}^{+} \cdot uc_{f,h}^{up} \quad \forall f, h \tag{31}$$

$$q_{f,h}^{dw} \leq Q_{f,h}^{-} \cdot uc_{f,h}^{dw} \quad \forall f, h \tag{32}$$

$$uc_{f,h}^{up} \cdot MinBidSize \leq q_{f,h}^{up} \quad \forall f, h \tag{33}$$

$$uc_{f,h}^{dw} \cdot MinBidSize \leq q_{f,h}^{dw} \quad \forall fh \tag{34}$$

$$uc_{f,h}^{up} + uc_{f,h}^{dw} \leq 1 \quad \forall f, h \tag{35}$$

$$P_{i,j}^{-} < p_{i,j,h} < P_{i,j}^{+} \quad \forall i, j, h \tag{36}$$

$$\theta_i^{-} < \theta_{i,h} < \theta_i^{+} \quad \forall i, h \tag{37}$$

$$q_{f,h} = q_{f,h}^{up} - q_{f,h}^{dw} \quad \forall f, h \tag{38}$$

$$Q_f^{-} < q_{f,h} < Q_f^{+} \quad \forall f, h \tag{39}$$

### C. Common CS with Separate CM and Balancing

This CS is characterized by a central MO, in this case, the TSO, solving congestions and balancing in sequential markets. Resources connected to the distribution and transmission grid are used for that purpose.

#### 1) Congestion Management Market

The congestion management market is characterized by a DC OPF, similar to the Common Joint CS, differing by the fact that the demand balance equations (40)-(41) do not include the imbalances.

Min (13)
s.t.
(16)-(39),

$$DispatchDA_{i,h} + \sum_{f \in IF} q_{f,h} - \sum_j p_{i,j,h} + \sum_j p_{j,i,h}$$
$$= D_{i,h} + nsf_{i,h}^p - nsf_{i,h}^n \quad \forall i \in IS, (s \in TS) \wedge (t = TSO), h \tag{40}$$

$$DispatchDA_{i,h} + \sum_{f \in IF} q_{f,h} - \sum_j p_{i,j,h} + \sum_j p_{j,i,h}$$
$$= D_{i,h} + nsf_{i,h}^p - nsf_{i,h}^n \quad \forall i \in IS, (s \in TS) \wedge (t = DSO), h \tag{41}$$

Following the congestion management market, minimums and maximums are adjusted and passed on to the balancing market, as exemplified in (42).

$$Q_{f,h}^{new+} = Q_{f,h}^{+} - q_{f,h}^{*} \quad \forall f, h \tag{42}$$

#### 2) Balancing Market

In the balancing phase, another DC OPF is run by the TSO, this time including only the imbalances in the demand balance equations (43)-(44) and considering the new limits received from the congestion management market.

Min (13)
s.t.
(16)-(39),

$$\sum_{f \in IF} q_{f,h} - \sum_j p_{i,j,h} + \sum_j p_{j,i,h} = \sum_{g \in IG} Imb_{g,h} \quad \forall i \in IS, (s \in TS) \wedge (t = TSO), h \tag{43}$$

$$\sum_{f \in IF} q_{f,h} - \sum_j p_{i,j,h} + \sum_j p_{j,i,h} = \sum_{g \in IG} Imb_{g,h} \quad \forall i \in IS, (s \in TS) \wedge (t = DSO), h \tag{44}$$

### D. Multi-level CS

In the Multi-level implementation, firstly, the DSO runs a local congestion management market, followed by the TSO market(s). For the Local Flexibility Markets (LFM), two different implementations are proposed. The first one is based on an OPF for the distribution network. The second LFM implementation is a reduction of the problem, considering only the power flow at the interface substations. The latter is a replication of a real LFM implemented in Sweden and discussed in [22].

#### 1) Local Flexibility Market – OPF

In the LFM, the DSO minimizes the cost of activating resources connected to the distribution grid to solve local congestions only. The demand balance equations for this market consider the results of the day-ahead market in terms of generation and demand for each node of the DSO's grid, plus the power expected at the interface with the TSO. In the Local Market of the Multi-level CS, the DSO expects to receive (or export, in the case of generation-driven grids) a certain power through the interfaces. In this case, the virtual demand $vd_{sh}$ variable is used. Equations (46)-(47) compute and allocate the power at the interfaces. Equations (48)-(49) calculate the subscription levels and the subscription costs incurred by the DSO, as in Sweden the DSO has a virtual power flow limit on the substation subject to potential penalties if surpassed. In general, congestions are characterized by the thermal limits of elements (given in (36)), but in Sweden, a "virtual congestion" set by regulation also applies. These formulations, however, can be used without compromise in power systems without subscription levels.

$$\min \sum_{s \in TS \wedge t=DSO, i, f, h} \left[ (Bid_f \cdot q_{f,h}^{up}) + (Bid_f \cdot q_{f,h}^{dw}) + subscost_{sh} + (nsf_{i,h}^{up} + nsf_{i,h}^{dw}) \cdot CNSF \right] \tag{45}$$

s.t.
(15), (17)-(39),

$$psubs_{i,h} = + \sum_{j \notin SUBS} p_{i,j,h} - \sum_{j \notin SUBS} p_{j,i,h} \quad \forall i \in (IS, SUBS), (s \in TS) \wedge (t = DSO), h \tag{46}$$

$$psubs_{i,h} = vd_{s,h} \cdot Impact_i \quad \forall i \in (IS, SUBS), (s \in TS) \wedge (t = DSO), h \tag{47}$$

$$psubs_{i,h} = \sum_{lv} qsubs_{i,lv,h} \quad \forall i \in (IS, SUBS), (s \in TS) \wedge (t = DSO), h \tag{48}$$

$$subscost_{s,h} = \sum_{i,lv} qsubs_{i,lv,h} \cdot CSubs_{i,lv} \quad \forall (s \in TS) \wedge (t = DSO), h \tag{49}$$

After the LFM, unused bids and the information on the activated FSPs are passed on to the TSO (in a similar fashion as in (42)). Unused bids, however, are passed on so that no congestions can be created. It is to say that the TSO cannot activate FSPs in the direction opposite to the direction they were activated by the DSO. Moreover, the information on the activated FSPs must be sent to the TSO (50), so this SO can consider it when forecasting the necessary power to be delivered at the interfaces (as in (53)).

$$DispatchFSP_{fh} = q^*_{fh} \quad \forall f, h \tag{50}$$

*2) Local Flexibility Market – PTDF-Based*

The second formulation of the LFM aims at replicating a Swedish implementation of the LFM. The DSO considers the impact of each demand, generation, and FSP activation over the substation using Power Transfer Distribution Factors (PTDFs). This computation is done in (51)-(53).

Min (45)
s.t.
(19)-(39), (48)-(49),

$$pfsub_{i \in SUBS, j, h} = \left( -\sum_{ii} DispatchDA_{ii,h} - \sum_{i,f} q_{f \in IF,h} + \sum_{ii} D_{ii,h} + \sum_{i} nsf^{up}_{ii,h} - \sum_{i} nsf^{dw}_{ii,h} \right) \cdot PTDF_{i,j,ii} \quad \forall i \in (IS, SUBS), j, (s \in TS) \wedge (t = DSO), h \tag{51}$$

$$psubs_{i,h} = \sum_j pfsub_{i \in SUBS, j, h} \quad \forall i \in (IS, SUBS), (s \in TS) \wedge (t = DSO), h \tag{52}$$

*3) TSO Market (Joint procurement)*

Following the DSO LFM, the TSO runs its market(s). In the following formulation, the joint balancing and congestion management market are described. Equation (53) computes the "virtual demand" that the TSO has to deliver to the DSO for the joint CS. Considering that we focus on HV distribution grids and those can be meshed, there could be multiple TSO-DSO interfaces for a single distribution grid. In this context, the virtual demand must be distributed to the different TSO-DSO interfaces. To do that, it is considered that the power flow on the interfaces will follow a typical impact factor for each substation. This impact factor is an average of the PTDFs of the nodes in the distribution grid in relation to the interface substations. This method intends to account for what, in reality, would be the forecasting process of the TSO. This process of allocating the virtual demand to the interface substations is done in (54)-(57).

The model below describes the joint TSO market in the Multi-level CS. The separate market follows the same logic as in the Common separate CS, in which the parameter $Imb_{gh}$ is not included in the supply-demand balance constraint (e.g. (53)), followed by the balancing market that includes $Imb_{gh}$ but not $DispatchDA_{ih}$ and $D_{ih}$, in a similar fashion as in (40) and (43).

Min (13)
s.t.
(14), (16), (19)-(39),

$$-\sum_i DispatchDA_{i,h} + \sum_i D_{i,h} + \sum_i Imb_{i,h} - \sum_{i \in IF} q_{f,h} - \sum_f DispatchFSP_{f,h} + \sum_i nsf^{up}_{i,h} - \sum_i nsf^{dw}_{i,h} = vd_{s,h} \quad \forall i \in IS, (s \in TS) \wedge (t = DSO), h \tag{53}$$

$$\sum_{j \in (L, SUBS)} p_{j,i,h} = pimport_{i,h} \quad \forall i \in (IS, SUBS), (s \in TS) \wedge (t = DSO), h \tag{54}$$

$$pimport_{i,h} = vd_{s,h} \cdot Impact_i \quad \forall i \in (IS, SUBS), (s \in TS) \wedge (t = DSO), h \tag{55}$$

$$pimport_{i,h} = pexport_{i,h} \quad \forall i \in SUBS, h \tag{56}$$

$$-\sum_j p_{i,j,h} + \sum_j p_{j,i,h} = pexport_{i,h} \quad \forall i \in (IS, SUBS), (s \in TS) \wedge (t = TSO), h \tag{57}$$

III. CASE STUDY

In Sweden, the regional DSO operates the subtransmission network and is already faced with the need for procuring distributed flexibility, as they are subject to subscription limitations at their interfaces with the TSO [22]. The subscription limits are virtual power flow limitations imposed at the substation. If surpassed, the DSO may incur a penalty. Hence, DSOs in Sweden are experimenting with procuring local flexibility as an alternative to potential penalties.

The models described above were run for a case study considering both a simplified transmission model for the whole country and a subtransmission model for the Uppsala region.

The Swedish transmission grid is an adapted version of the Nordic 32 [23]. This 32-node test system is a fictitious but similar grid to the Swedish one, with connections to the Nordic system. Additional modifications and inclusions were necessary in order to make this test system a more robust representation of the current Swedish system. For this purpose, [24], [25] were used.

The networks based on the Nordic 32 test system, however, do not include the distribution network necessary for the TSO-DSO study. For that purpose, a sub-transmission grid is incorporated into the transmission network. This sub-transmission grid is a representation of the 70 kV network of Uppsala city, one of the demonstration sites of the Swedish demonstration in the H2020 CoordiNet project [26]. Fig. 2 provides an illustration of the distribution grid considered.

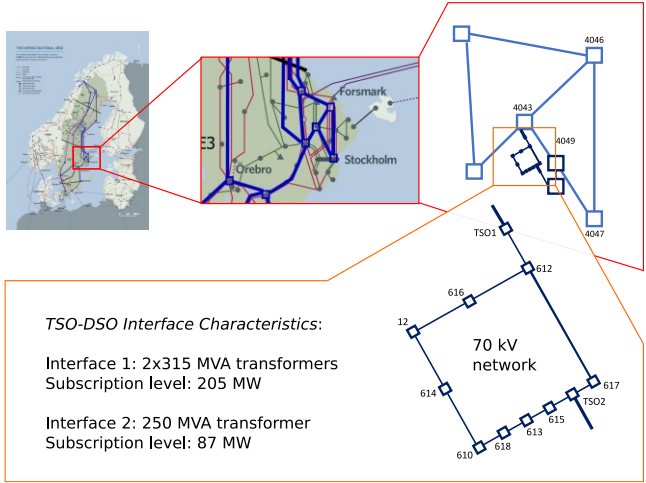

**Fig. 2.** Location and representation of the 70 kV Uppsala grid.

With regard to the load and generation parameters, 2020 is used as the base year. The load profiles for the whole year are gathered and clustered into eight representative days by the employment of the k-means clustering method. Two representative days are calculated per season, representing high and low loads (weekdays and weekends). The data used was gathered from the Swedish TSO Svenska kraftnät's website.

The generation of renewables is aggregated into clusters per bidding zone. One relevant characteristic of the Swedish wholesale market is the existence of four different bidding zones. The Net transfer Capacity (NTC) between bidding zones considered is the maximum NTC calculated by ENTSO-e and published by Nord Pool [27].

The cost information per type of technology considered is obtained from [28]. For wind farms and solar power plants, the variable cost considered is close to 1 €/MWh, as no fuel costs exist. Additionally, the cycling costs from [28] are also used. It is important to highlight that the objective of the analyses in this modelling workstream is not to forecast actual costs for SOs, but rather to study the rate of change under the different scalability and replicability scenarios.

### A. Results from the Day-ahead market

The results from the day-ahead market will determine the congestion needs later considered in the individual CSs. For this reason, the obtained results from the DA market for one year are compared with the actual DA market results in terms of the energy mix and average price. Technologies such as nuclear and thermal generators will have a minimum dispatch associated with them, also obtained from [28]. For nuclear technology only, a "must-run" option is added, considering that the model runs eight individual representative days without constraints connecting them. It is observed that the model reaches satisfactory comparability with the actual generation in Sweden in the year 2020, as shown in Table I. The difference in total energy generated by the model (141 TWh) against the actual generation (152 TWh) is due to the higher export of energy to the connecting systems, which are not fully modeled.

TABLE I
COMPARISON BETWEEN MODEL OUTPUT AND ACTUAL GENERATION MIX IN SWEDEN IN 2020

| Technology | Model Output | | Actual Gen. Mix | |
|---|---|---|---|---|
| | TWh | % Mix | TWh | % Mix |
| Thermal | 3 | 2% | 6 | 4% |
| Hydro | 68 | 48% | 72 | 47% |
| Nuclear | 43 | 31% | 47 | 31% |
| Wind | 27 | 19% | 28 | 18% |
| Total | 141 TWh | | 153 TWh | |

In order to evaluate the different CSs, appropriate scenarios of congestion management and balancing needs are required. As mentioned above, the overall need for congestion management is generated by the DA dispatch and the network capability. The congestions from the DA market are the same in every CS.

According to [29], the total volume of remedial actions in 2020 in Sweden was 69.2 GWh, and the total cost was 1.14 million euros. These values are taken as a reference. The balancing needs are included as an input of the model assigned per node of the grid based on the results of the DA clearing. In order to calculate the amount of balancing needs and in which direction, data from the ENTSO-E transparency platform is used [30]. It is observed that imbalances in Sweden totaled approximately 3 TWh in 2020. Imbalances are equally distributed among deficit and surplus hours.

### B. Scenarios

#### 1) Base Case

The flexibility available for the different market sessions in the different CSs is offered by resources connected at both the transmission and the distribution grids. The FSPs connected to the transmission grid are the same generators that participate in the DA market. Their capability to offer flexibility will depend on their cleared volume in the DA market and their technology. We assume that wind farms can only offer 5% of their DA schedule upwards, considering that there are differences between their DA forecasts and the real-time generation [31], [32]. Solar power plants cannot offer upward flexibility, only downward. In terms of bidding, these units bid the same variable cost as they offered to the DA market. In this context, we assume that the FSPs are participating under perfect competition and act as naïve agents and therefore bid their true short-term variable cost, considering the scope of this paper.

In the Swedish case, the FSPs connected to the distribution grid are based on the FSPs participating in the demonstration activities of the CoordiNet project [33]. Table II lists the type of FSP and their capacity to offer upward and downward flexibility. One battery is included, with an energy capacity of 20 MWh and a flexibility capacity of 5 MW, having to comply with the state of charge (SoC) formulations (24)-(30). The round-trip efficiency considered is 80% [34].

TABLE II
FSPs CONNECTED TO THE UPPSALA GRID IN SWEDEN

| FSP | BUS | FSP type | Downward flexibility (MW) | Upward flexibility (MW) | Bid prices (€/MWh) |
|---|---|---|---|---|---|
| 1 | 612 | Battery (20 MWh) | 5 | 5 | 8 |
| 2 | 613 | Office buildings | 0 | 0.5 | 10 |
| 3 | 12 | Multi-family housings | 0 | 0.5 | 16 |
| 4 | 610 | Commercial building | 0 | 0.5 | 12 |
| 5 | 613 | District heating | 5 | 30 | 20 |
| 6 | 613 | Multi-family housings | 0 | 0.5 | 16 |
| 7 | 613 | Industry | 0.5 | 1 | 16 |
| 8 | 614 | Industry | 0.5 | 1 | 16 |

The base case for the scalability and replicability scenarios is the "Multi-level (PTDF-based)" CS considering the FSPs connected at the distribution grid.

*2) Sensitivities*

For the scalability and replicability analysis, different scenarios are tested. For scalability purposes, a sensitivity analysis is used. Different sensitivities are run in order to evaluate the effects on increased demand, increase in available distributed flexibility, and changes in flexibility bids. Sensitivity factors are applied to selected parameters of the optimization models, as presented in Table III. The sensitivity range shows the values to which parameters are multiplied in the sensitivity analysis.

TABLE III
SENSITIVITY FACTORS FOR SCALABILITY ANALYSIS

| Parameter | Considerations | Sensitivity range |
|---|---|---|
| $Q_f^+, Q_f^-$ | Applied only to FSPs connected at the distribution grid | [0  0.2  ⋯  2.8  3] |
| $Bid_f$ | Applied only to FSPs connected at the distribution grid | [0  0.2  ⋯  2.8  3] |
| $D_{ih}$ | Applied only to the load connected at the distribution grid | [0.8  0.9  ⋯  1.9  2] |

The replication scenarios considered for the Swedish case are two, namely, the different CSs and the types of FSPs. The latter proposes an exercise in connecting different types of FSPs not observed in the Swedish network being analyzed. In this case, two wind farms are connected to the distribution grid.

IV. RESULTS

Table IV presents the results in terms of energy activated for the base case scenario. This base case scenario considers the "Multi-level (PTDF-based)" CS with the FSPs connected to the distribution grid. Values presented are in GWh/year, as results per representative day are already multiplied by the number of times the representative day takes place in one year. It is possible to observe that activations due to congestion management needs do occur, but only during the winter representative days. The DSO does activate 10 GWh of flexibility in its LFM, which represents 0.69% of the total energy supplied by the distribution grid. The TSO activates approximately 3.6 TWh, of which 3 TWh are due to balancing needs and 0.6 are due to congestions.

TABLE IV
BASE CASE SCENARIO FOR SWEDISH CASE STUDY: ENERGY ACTIVATED. IN GWh/YEAR.

| SO Market Product Direction | Winter | | Spring | | Summer | | Autumn | | Total Year |
|---|---|---|---|---|---|---|---|---|---|
| | High | Low | High | Low | High | Low | High | Low | |
| DSO LFM | **9** | **1** | **0** | **0** | **0** | **0** | **0** | **0** | **10** |
| *CM* | *9* | *1* | *0* | *0* | *0* | *0* | *0* | *0* | *10* |
| Up. | 9 | 1 | 0 | 0 | 0 | 0 | 0 | 0 | 10 |
| TSO Markets | **1,008** | **368** | **338** | **427** | **438** | **320** | **326** | **420** | **3,645** |
| *B* | *400* | *341* | *337* | *427* | *438* | *320* | *326* | *420* | *3,009* |
| Down. | 189 | 175 | 132 | 98 | 153 | 138 | 154 | 164 | 1,203 |
| Up. | 211 | 166 | 206 | 329 | 285 | 182 | 172 | 256 | 1,807 |
| *CM* | *608* | *27* | *0* | *0* | *0* | *0* | *0* | *0* | *635* |
| Down. | 308 | 14 | 0 | 0 | 0 | 0 | 0 | 0 | 323 |
| Up. | 300 | 13 | 0 | 0 | 0 | 0 | 0 | 0 | 313 |

When comparing the total cost for a scenario with no flexibility being provided by the FSPs at the distribution grid with the base case scenario, it is possible to verify a significant reduction for the DSO from 1,674 k€ per year to 802 k€. In a scenario with no flexibility, the totality of the cost would be related to the potential penalties associated to the surpassing of subscription levels. Table V presents the results for both scenarios, as well as the replication scenarios comparing different CSs.

TABLE V
COMPARISON BETWEEN THE "NO-FLEXIBILITY" AND THE "BASE CASE" SCENARIOS IN SWEDEN. IN k€/YEAR.

| Market Model | No-Flexibility Scenario | Base Case Scenario |
|---|---|---|
| **Common** | | |
| Joint | 13,095 | 11,658 |
| Separate | 13,793 | 12,481 |
| **Multi-level (OPF)** | | |
| Local | 1,674 | 713 |
| Joint | 11,913 | 11,937 |
| Separate | 12,751 | 12,767 |
| **Multi-level (PTDF-Based)** | | |
| Local | 1,817 | 802 |
| Joint | 11,913 | 11,937 |
| Separate | 12,751 | 12,769 |

When comparing the results from the different CSs, it is possible to observe that the common market model would lead to the least total cost of flexibility procurement, considering that for both implementations of the multi-level CS, the LFM cost had to be added to either the joint or separate TSO markets in order to make costs comparable. It is also noticeable that the subscription limitation works as a transferring mechanism of congestion costs from the TSO to the DSO, as observed in the No-Flexibility Scenario. It showcases the use of virtual congestions and financial incentives as a way of congestion management cost allocation.

Fig. 3 below presents the results for the scalability scenario in which sensitivity factors are applied to the sizes and bids of the FSPs connected to the distribution network. The costs shown are for the DSO in the LFM. The sensitivity factor in the x and y axes of the graph are those defined in Table III, where a sensitivity factor of 1 represents the base case scenario. In the curve of the graph, it is possible to identify the "no-flexibility" and the "base case" scenarios discussed above and presented in Table V. Moreover, this scalability scenario reveals that if the capacity of the FSPs in the demo is scaled up to a factor of 1.6, subscription penalties are eliminated by the procurement of local flexibility.

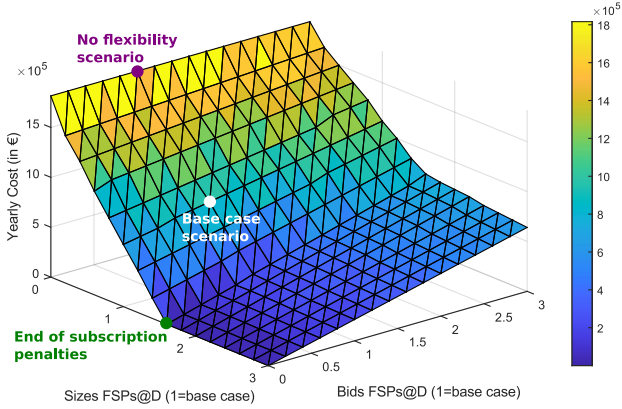

**Fig. 3.** Sensitivities to size and bids of FSPs connected at distribution (FSPs@D). DSO costs in the Multi-level (PTDF-Based) LFM.

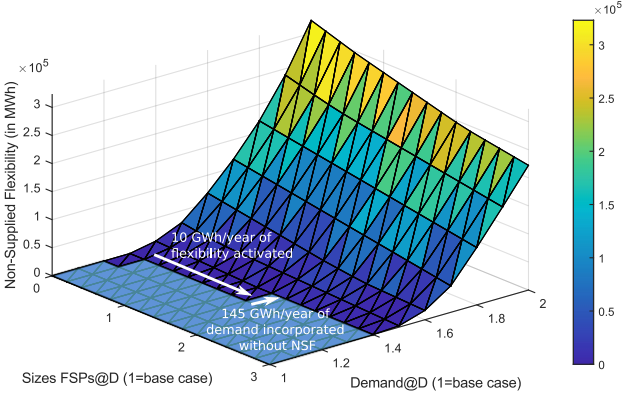

**Fig. 4.** Sensitivity to demand connected to distribution (Demand@D) and size of FSPs connected at distribution (FSPs@D). Non-Supplied Flexibility for DSO in Multi-level (OPF) LFM.

The second scalability scenario explores the effects of increasing the demand connected to the distribution grid. In this scenario, we run a sensitivity factor over the demand and the size of FSPs in the distribution grid and explore the concept of "non-supplied flexibility" (NSF). The NFS is the position in which the DSO has congestions in its grid and wants to procure flexibility to solve these congestions, but the flexibility available in the market is not sufficient or not effective for that purpose. In that case, the DSO would have to resort to mechanisms other than the flexibility market to ensure the secure operation of the grid (e.g. change in topology, curtailment of selected units). Fig. 4 presents the result of this scalability scenario.

The flat light-blue area on the graph represents the region in which either the DSO does not have any congestion in the network or, if congestions exist, they can be solved by the available flexibility in the LFM. Outside the flat area is the region in which the DSO observes some amount of non-supplied flexibility. It is important to note, though, that this analysis considers only congestions due to thermal limit violations and not needs due to subscription penalties (which already exist at the current load level). For this reason, this scalability scenario considers the CS Multi-level (OPF), which also accounts for the power flow in every line in the grid, and not only the power flow at the substation as the Multi-level (PTDF-Based).

This analysis reveals that an increase in the available FSPs capacity from 0.2 to 1.2 would allow the DSO to incorporate another 10% of demand without entering into the non-supplied flexibility region. In energy terms, this means that an increase of 10 GWh of activated flexibility per year would allow the incorporation of 145 GWh of demand without leading to an NSF situation. For the DSO, this could mean that grid reinforcement needed in the face of demand growth could be deferred using local flexibility, for instance.

Finally, the replication case is presented, in which two wind farms are incorporated into the Swedish case study. The addition of the two wind farms consistently reduces the total costs for the TSO by approximately 2%. For the DSO, however, the costs can be reduced by up to 98%. This is since incorporating the two new DERs in the DSO's grid not only increases the flexibility available but also means that distributed generation is included in this grid. The DG helps to offset the need for importing energy from the TSO through the TSO-DSO interfaces, which leads to a significant reduction in subscription penalties. Table VI presents the results from the second replication scenario.

TABLE VI
REPLICATION SCENARIO. WIND FARMS INCORPORATED TO SWEDEN CASE STUDY. IN k€/YEAR.

| Coordination Scheme | Base Case | Replication Scenario | % |
|---|---|---|---|
| **Common** | | | |
| Joint | 11,658 | 11,369 | -2.5% |
| Separate | 12,481 | 12,197 | -2.3% |
| **Multi-level (OPF)** | | | |
| Local | 713 | 12 | -98.3% |
| Joint | 11,937 | 11,674 | -2.2% |
| Separate | 12,767 | 12,514 | -2.0% |
| **Multi-level (PTDF-Based)** | | | |
| Local | 802 | 23 | -97.2% |
| Joint | 11,937 | 11,683 | -2.1% |
| Separate | 12,769 | 12,521 | -1.9% |

## V. CONCLUSION

In this paper, a techno-economic dispatch model is proposed for the evaluation of different TSO-DSO CSs in the context of balancing and congestion management services being provided by DERs and considering the transmission network and the HV distribution network in a meshed-to-meshed topology. This model is then applied to realistic case study on the Swedish power system, including the modeling of one year based on real demand and generation data.

The comparison of the different CSs shows that the Common CS leads to the overall least cost of flexibility procurement. The Multi-level CS with subscription penalties showcased the use of virtual congestions and financial incentives as a way of congestion management cost allocation.

Once local flexibility is used at the distribution grid, however, the TSO starts observing a higher cost for congestion management and balancing, as the access to distributed flexibility in a Multi-level CS might be sub-optimal.

From the DSO's perspective, it is shown that, for the analyzed case study, an increase of only 60% over the base case flexibility could already lead to a situation in which the DSO does not incur subscription penalties.

Considering that the grid studied was load-driven, the increase in FSPs availability could also be beneficial when coping with the increase in demand. The study suggested that an increase of 10 GWh of activated flexibility per year would allow the incorporation of 145 GWh of demand without leading to an NSF situation for the DSO.

The replication scenario in which wind farms are incorporated into the distribution grid shows that this type of DER could help the DSO to mitigate the surpassing of subscription levels, as the incorporation of distributed generation offsets the need for imports from the TSO through the TSO-DSO interface.

## VI. Acknowledgment

The authors would like to thank the partners of the Swedish Vattenfall Distribution company for the inputs and comments provided. This work is supported by the European Union Horizon 2020 research and innovation program under grant agreement No. 824414 – CoordiNet Project.